\documentclass[twocolumn, superscriptaddress,preprintnumbers,amsmath,amssymb,longbibliography]{revtex4-1}
\usepackage{graphicx}% Include figure files
\usepackage{dcolumn}% Align table columns on decimal point
\usepackage{bm}% bold math
\usepackage{mathtools}
\usepackage{microtype}
\usepackage{color,soul}
\usepackage{amsmath}
\usepackage{csquotes}
\usepackage{setspace}

\begin{document}

\preprint{Version: \today}

\title{Tunable Broadband Transparency of Macroscopic Quantum Superconducting Metamaterials}

\author{Daimeng Zhang}
\email{dmchang@umd.edu}
\affiliation{Department of Electrical and Computer Engineering, University of Maryland, College Park, Maryland 20742-3285, USA}
\affiliation{Center for Nanophysics and Advanced Materials, Department of Physics, University of Maryland, College Park, Maryland 20742-4111, USA}
\author{Melissa Trepanier}
\affiliation{Center for Nanophysics and Advanced Materials, Department of Physics, University of Maryland, College Park, Maryland 20742-4111, USA}
\author{Oleg Mukhanov}
\affiliation{Hypres, Inc., 175 Clearbrook Road, Elmsford, New York 10523, USA}
\author{Steven M. Anlage}
\affiliation{Department of Electrical and Computer Engineering, University of Maryland, College Park, Maryland 20742-3285, USA}
\affiliation{Center for Nanophysics and Advanced Materials, Department of Physics, University of Maryland, College Park, Maryland 20742-4111, USA}

\date{\today}% It is always \today, today,

\begin{abstract}

Narrow-band invisibility in an otherwise opaque medium has been achieved by electromagnetically induced transparency (EIT) in atomic systems . The quantum EIT behaviour can be classically mimicked by specially engineered metamaterials via carefully controlled interference with a "dark mode" .  However, the narrow transparency window limits the potential applications that require a tunable wide-band transparent performance. Here, we present a macroscopic quantum superconducting metamaterial with manipulative self-induced broadband transparency due to a qualitatively novel nonlinear mechanism that is different from conventional EIT or its classical analogs. A near complete disappearance of resonant absorption under a range of applied rf flux is observed experimentally and explained theoretically. The transparency comes from the intrinsic bi-stability of the meta-atoms and can be tuned on/ off easily by altering rf and dc magnetic fields, temperature and history. Hysteretic $in$ $situ$ $100\%$ tunability of transparency paves the way for auto-cloaking metamaterials, intensity dependent filters, and fast-tunable power limiters.

\end{abstract}

\maketitle

\section{Introduction}
Controllable transparency in an originally opaque medium has been an actively studied topic. Among those efforts, electromagnetically induced transparency (EIT) in three-level atomic systems is one of the most compelling ideas \cite{Harris1990,Harris1997,EITreview3}. A classical analog of EIT atomic systems is realized in structured metamaterials that have a radiative mode coupled to a trapped mode \cite{Fedotov2007, Kurter2011}. The interaction between two modes induces a narrow transparency window in which light propagates with low absorption, also creating strong dispersion and a substantial slowing of light \cite{Harris1990, Harris1997, EITreview3, Fedotov2007, Kurter2011}. Superconducting metamaterials, especially, exhibit intrinsic nonlinearity thus opening the door for tunable electromagnetic transparency \cite{Abdumalikov2010,Fedotov2010,Jung2014review,Kurter2012,Tsironis2014}. Nevertheless, most techniques that induce transparency require careful manipulation of coupling, and only allow light to propagate with perfect transmission in a narrow frequency bandwidth inside a broad resonance feature. The narrow-band invisibility prevents conventionally induced transparency from applications requiring broad-band invisibility. In this work, we demonstrate a microwave metamaterial with self-induced broadband and tunable transparency that arises from an altogether new mechanism. 

Our macroscopic quantum superconducting metamaterial is made of Radio Frequency Superconducting QUantum Interference Device (rf-SQUID) meta-atoms. An rf-SQUID is a macroscopic quantum version of the split ring resonator in that the gap capacitance is replaced with a Josephson junction. The rf-SQUID is sensitive to the applied rf and dc magnetic flux, and the scale of this response is the flux quantum $\Phi_0=h/2e=2.07\times10^{-15}$ Tm$^2$, where $h$ is Planck's constant and $e$ is the elementary charge. The rf-SQUID combines two macroscopic quantum phenomena: magnetic flux quantization and the Josephson effect \cite{Tinkham1996}, making it extremely nonlinear and tunable \cite{Anlage2011, Jung2014review, Lapine2014}. Incorporating rf-SQUIDs into metamaterials has received increasing attention \cite{Du2006, Lazarides2007,Maimistov2010, Lazarides2013,Jung2013,Butz20132,Trepanier2013, Jung2014}. The large-range resonance tunability by dc magnetic flux at low drive amplitude \cite{Jung2013,Butz20132,Trepanier2013,Jung2014}, and switchable multistability at high drive amplitude \cite{Jung2014} of rf-SQUID metamaterials have been studied experimentally, however, the broadband response of the rf-SQUID metamaterial under intermediate rf flux range has not been systematically examined yet. 

Here we demonstrate that in the intermediate rf flux range the rf-SQUID metamaterial develops self-induced broadband and tunable transparency that arises from the intrinsic nonlinearity of rf-SQUIDs. Both experiment and simulation show that the resonance of this metamaterial largely disappears when illuminated with electromagnetic waves of certain power ranges. We can adroitly control the metamaterial to be transparent or opaque depending on stimulus scanning directions. The degree of transparency can be tuned by temperature, rf and dc magnetic field. We also discuss analytical and numerical models that reveal the origin of the effect and how to systematically control the transparency regime. The nonlinear transparency behaviour should also extend to the quantum regime of superconducting quantum metamaterials interacting with a small number of photons \cite{Bishop2010,Abdumalikov2010,Boissonneault2010,Reed2010, Macha2014}. The observed tunable transparency of the rf SQUID metamaterial offers a range of previously unattainable functionalities because it acts effectively as a three-terminal device. New applications include wide-band power limiters for direct-digitizing rf receivers \cite{Mukhanov2008}, gain-modulated antennas \cite{Mukhanov2014}, rf pulse shaping for qubit manipulation,  tunable intensity-limiting filters, and the novel concept of a metamaterial that spontaneously reduces its scattering cross-section to near zero (auto-cloaking), depending upon stimulus conditions.

\begin{figure}[]
\includegraphics[width=80 mm]{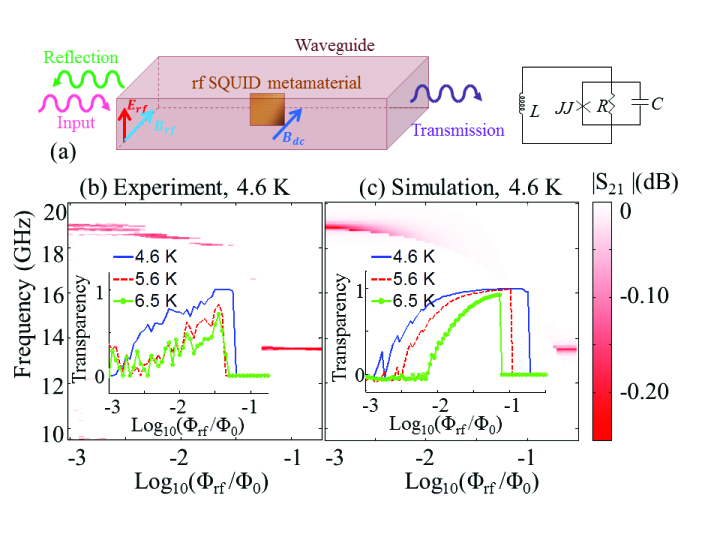}
\caption{(a) Schematic of the experimental setup for measuring the transmission of rf-SQUID metamaterial oriented perpendicular to rf magnetic field inside a rectangular waveguide. The dc magnetic field is applied by sending current into a superconducting coil outside the waveguide. The inset of (a) is the Resistively and Capacitively Shunted Junction (RCSJ) model of an rf-SQUID. Transmission of a single meta-atom at $4.6$ K depending on frequency and rf flux is shown in (b) experiment and (c) simulation. Red (dark) feature denotes the resonance dip. Insets plot the transparency as a function of input power under 4.6 K (solid blue), 5.6 K (dashed red) and 6.5 K (dot line green) both for (b) experiment and (c) simulation.}
\label{fig1}
\end{figure}

\section{Results}
\subsection{Broadband transparency at a fixed applied dc flux}
We measure the transmission and reflection of our metamaterial samples positioned in a rectangular waveguide so that the rf magnetic field of the propagating wave is perpendicular to the SQUID loop (Fig. \ref{fig1} (a)) (see Supplemental Material) \cite{Trepanier2013}. Measured transmission of a single meta-atom as a function of frequency and rf flux $\Phi_{rf}$ at a temperature of $4.6$ K under $0$ dc flux is shown in Fig. \ref{fig1} (b). Red features denote the resonance absorption dips of the meta-atom. At low input rf flux, the resonance is strong at $19$ GHz \cite{Trepanier2013}. In the intermediate rf flux range, the resonance shifts to lower frequency and systematically fades away ($|S_{21}|=0$ dB) through the entire frequency range of single-mode propagation through the waveguide. At an upper critical rf flux, a strong resonance abruptly appears at the geometrical resonance frequency $\omega_{geo}/2\pi=13.52$ GHz for a single rf-SQUID. We employ the nonlinear dynamics of an rf-SQUID to numerically calculate transmission (see Supplemental Material) shown in Fig. \ref{fig1} (c) which shows the same transparency behaviour. 

We define a normalized transparency level that quantitatively determines the degree of resonance absorption compared to the low rf flux absorption (see Supplemental Material). High transparency indicates a weak resonance absorption. The extracted transparency shows a clear onset rf flux for transparency and an upper critical rf flux determining the abrupt end of transparency (insets of Fig. \ref{fig1} (b) and (c)). The transparency approaches $1.0$ between these rf flux values. The measurements are taken at $4.6$ K, $5.6$ K, and $6.5$ K; at lower temperature, both experiment and simulation show a larger range of transparency as well as a higher degree of transparency. 

\begin{figure}[]
\includegraphics[width=80mm]{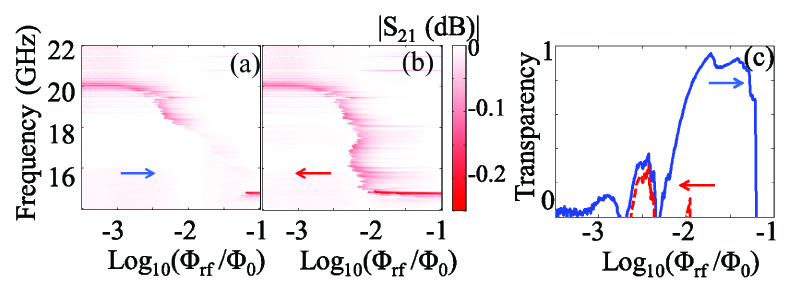}
\caption{Measured transmission as a function of frequency and applied rf flux for an 11x11 array rf-SQUID metamaterial, when the input rf flux (a) keeps increasing, and (b) continuously decreases. (c) Extracted Transparency values for sweep directions (a) and (b). The arrows denote the rf flux scanning directions. The temperature is $4.6$ K.}
\label{arraydata}
\end{figure}

\subsection{Hysteretic transparency}
Collecting rf-SQUID meta-atoms into a metamaterial preserves the self-induced broadband transparency performance. Fig. \ref{arraydata} illustrates the transmission of an $11\times11$ rf SQUID array metamaterial (see Supplemental Material for parameters) with interactions among the meta-atoms. The metamaterial is stimulated at fixed frequency while the rf flux amplitude is scanned under nominally $0$ applied dc flux at $4.6$ K. 
The resonance is almost invisible as the input rf flux increases continuously through the transparency range (Fig. \ref{arraydata} (a)). 

However, a reverse rf flux scan renders an opaque behaviour (Fig. \ref{arraydata} (b)): the resonance is strong across all rf flux values. Quantitatively, the transparency value reaches $0.9$ in the forward sweep and is below $0.3$ for the backward sweep (Fig. \ref{arraydata} (c)). We did numerical simulations on this metamaterial and they show the same hysteretic transparent/opaque behaviour. Similar hysteresis is also observed for measurements and simulations of a single rf-SQUID meta-atom. These observations mean that transparency can be turned on and off depending on the stimulus scan direction and metamaterial history. 

\subsection{Transparency and bi-stability}
\begin{figure}[]
\includegraphics[width=80mm]{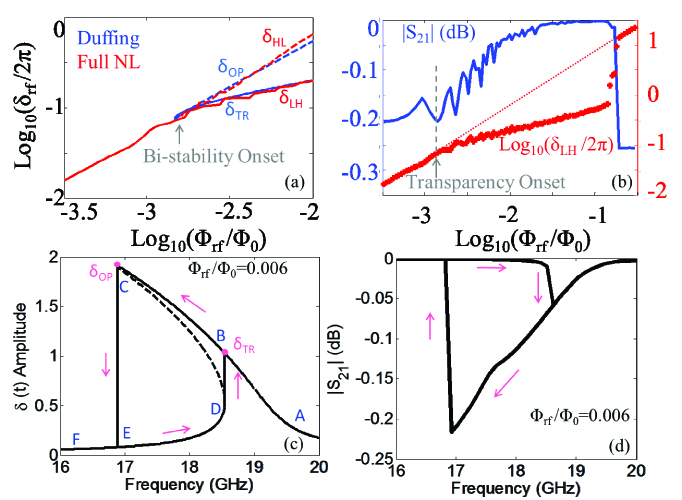}
\caption{Simulation results for a single rf-SQUID in the bi-stable regime. The pink arrows denote the signal (frequency or rf flux amplitude) scanning directions. (a) The amplitude of gauge invariant phase $\delta (t)$ oscillation on resonance as a function of driving rf flux amplitude. Red curves denote the full nonlinear numerical results. The $\delta_{LH}$ is the amplitude of $\delta (t)$ when the frequency or rf flux scans from low value to high value, and $\delta_{HL}$ is the amplitude for a reverse scan. Blue curves are calculated analytically with the Duffing oscillator approximation. The $\delta_{OP}$ and $\delta_{TR}$ denote the analytic amplitudes for the opaque state and the transparent state respectively. The gray arrow points out the onset of bi-stability. (b) The simulated transmission (blue curve, left y-axis) and the amplitude of $\delta (t)$ (red curve, right y-axis) on resonance as a function of rf flux for the transparent case. The red dashed line shows the $\delta(t)$ amplitude interpolated between the low-power and high-power limits. The gray arrow shows the onset rf flux for transparency. It is the same as the bi-stability onset. (c) The fold-over resonance of the amplitude of $\delta(t)$ calculated in the Duffing oscillator approximation results in a bistable oscillation on resonance with amplitudes $\delta_{TR}$ and $\delta_{OP}$ respectively at an rf flux of $0.006\Phi_0$. The dashed curve plots the unstable solutions in the fold-over resonance. (d) The numerically calculated transmission as a function of rf driving frequency at an rf flux of $0.006\Phi_0$ shows that the frequency range of bistability is the same as analytically predicted in Duffing oscillator approximation in (c).}
\label{mechanism}
\end{figure}

The origin of the nonlinear transparency is the intrinsic bi-stability of the rf-SQUID. The gauge invariant phase difference of the macroscopic quantum wavefunction across the Josephson junction, $\delta (t)$, and its time dependence, determine essentially all properties of the rf SQUID and the associated metamaterial (see Supplemental Material). In simulation we found that $\delta (t)$ is almost purely sinusoidal; the amplitudes of the higher harmonics are less than $1\%$ of the fundamental resonance amplitude for all drive amplitudes considered here. The amplitude of the gauge-invariant phase oscillation on resonance as a function of rf flux for a forward stimulus sweep ($\delta_{LH}$) is lower than the amplitude for a reverse sweep ($\delta_{HL}$) above the onset of bi-stability (Fig. \ref{mechanism} (a)). The lower gauge-invariant phase amplitude results in a smaller magnetic susceptibility and thus a reduction of resonant absorption. The relation between the resonance strength (degree of transparency) and $\delta_{LH}$ is shown in Fig. \ref{mechanism} (b) for the transparent state. The onset rf flux of transparency coincides with the abrupt reduction of $\delta_{LH}$-$\Phi_{rf}$ slope and the onset of bi-stability. 

We can apply the Duffing oscillator approximation to analytically predict the onset of bi-stability. For intermediate drive amplitude an rf-SQUID can be treated as a Duffing Oscillator (Kerr Oscillator) \cite{Bishop2010}, which is a model widely adopted for studying Josephson parametric amplifiers \cite{Vijay2009, Siddiqi2013}. This approximation suggests that when the drive amplitude reaches a critical value, the amplitude of the $\delta(t)$ oscillation as a function of frequency is a fold-over resonance (Fig. \ref{mechanism} (c)), creating bi-stable oscillating states (see Supplemental Material). The amplitudes of the gauge-invariant phase difference oscillation for the transparent state ($\delta_{TR}$) and the opaque state ($\delta_{OP}$) are calculated analytically for each rf flux, and compared to the amplitudes $\delta_{LH}$ and $\delta_{HL}$ calculated numerically  (Fig. \ref{mechanism} (a)). Both the onset of bi-stability and the amplitudes of the two states match very well. The bi-stability of $\delta(t)$ amplitudes explains the bi-stability of transmission observed in experiment and simulation (Fig. \ref{mechanism} (d)). The very good agreement between the Duffing oscillator and the full nonlinear numerical simulation allows us to study analytically how to enhance the transparency values and the transparency range.

The onset rf flux value for transparency depends on several parameters. Higher resistance in the junction, higher capacitance, and higher critical current all give a lower onset rf flux for transparency (Supplemental Material). Operating the metamaterial at a lower temperature increases the resistance and the critical current, thus decreasing the onset, explaining the modulation of onset by temperature observed in experiment and simulation (insets of Fig. \ref{fig1}). The applied dc flux has a more modest effect on the onset rf flux. With a dc flux of a quarter flux quantum, the single rf-SQUID has an onset that is $13\%$ smaller than the 0-flux case (see Supplemental Material).

\begin{figure}[]
\includegraphics[width=80mm]{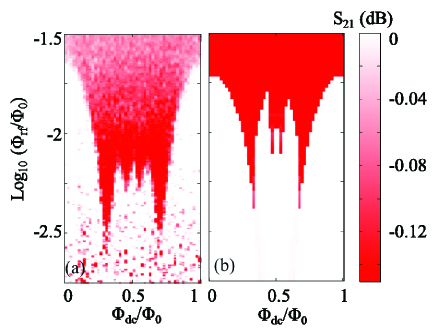}
\caption{The transmission as a function of rf flux and dc flux at the geometrical frequency $\omega_{geo}/2\pi$ of (a) experiment and (b) simulation for a single rf-SQUID at $4.6$ K. The strong resonance absorption (red region) at the geometrical frequency determines the upper critical rf flux of transparency. The edge between the red region and the white region denotes the tuning of upper critical rf flux by applied dc flux. The experiment is taken with pulsed rf measurement with a duty cycle of $1\%$ (see Supplemental Material). The pulse width is 2 $\mu$s which is long enough to ensures that the rf-SQUID achieves a steady state.}
\label{tunable_edge}
\end{figure}

\subsection{Modulation of transparency by dc flux}
The dc flux has a strong modulation of the transparency upper critical rf flux. Above the upper critical rf flux, the rf-SQUID experiences phase slips on each rf cycle and shows strong resonant absorption at the geometrical frequency $\omega_{geo}/2\pi$ \cite{Jung2014}. We can determine the transparency upper critical edge when the driving frequency is fixed at the geometrical frequency $\omega_{geo}/2\pi$, while rf flux amplitude scans from $0.001\Phi_0$ to $0.1\Phi_0$. The sudden decrease of transmission denotes the upper critical rf flux in differing amounts of dc flux (Fig. \ref{tunable_edge} (a)). The numerical simulation is depicted in Fig. \ref{tunable_edge} (b). There is a tunability of over a factor of $3.5$ in transparency upper critical flux by varying dc flux through the sample. Note that the entire dc magnetic field variation in Fig. \ref{tunable_edge} is only $10$ nT. The result shows that at a fixed frequency $\omega_{geo}/2\pi$ the meta-atom can be transparent or opaque depending very sensitively on the rf flux and dc flux.

\begin{figure}[]
\includegraphics[width=80mm]{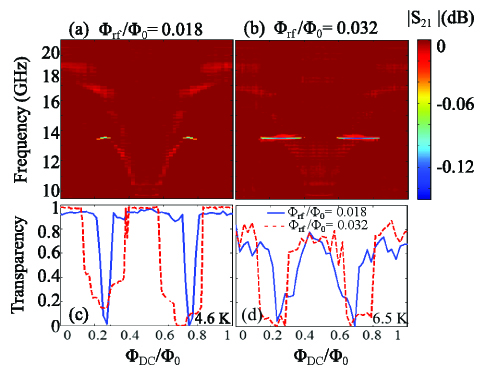}
\caption{The measured transmission at $4.6$ K of a single rf-SQUID meta-atom as a function of dc flux and frequency at different drive amplitudes: (a) rf flux$=0.018\Phi_0$, (b) rf flux$=0.032\Phi_0$. Dark red is the perfect transmission background, light red and blue features denote the weak and strong resonance absorption, respectively. The extracted transparency values for rf flux amplitude of $0.018\Phi_0$ (red dashed line) and $0.032\Phi_0$ (blue solid line) are shown for (c) $4.6$ K and (d) $6.5$ K. They both show a clear switchable transparent behaviour by dc flux.}
\label{dcflux}
\end{figure}

Also, at a fixed rf flux near the upper critical edge the sample can be resonantly absorbing at $\omega_{geo}$ or be transparent in the broadband frequency window depending sensitively on the applied dc flux. Fig. \ref{dcflux} plots the experimental transmission as a function of dc flux and frequency when our sample is illuminated with an rf flux amplitude of $0.018\Phi_0$ and $0.032\Phi_0$ respectively. Strong absorption at the geometrical resonance appears around $\Phi_0/4$ and $3\Phi_0/4$ dc flux values, while being broad-band transparent near 0 and 0.5 $\Phi_0$. Fig. \ref{dcflux} (c) shows that higher rf flux values pushes the rf-SQUID to be opaque under a larger range of dc flux, but the maximum value of transparency is higher than the lower rf flux case (Fig. \ref{dcflux} (c)). As the temperature increases, the transparency is weaker, as seen in Fig. \ref{dcflux} (d). The tuning of transparency by dc flux at a fixed rf flux indicates again a switchable on/off transparency behaviour with small variations of dc flux for our meta-atom.  

\section{Discussion}

Up to this point we have mainly discussed the results for a single rf-SQUID, because the transparency behaviour of the rf-SQUID metamaterial arises from the bi-stability of single meta-atoms. Disorder in the rf or dc flux can affect the degree of transparency in an rf-SQUID metamaterial but the effect is quite small. In experiments on an $11\times11$ array metamaterial, an intentionally introduced gradient of $0.11\Phi_0$ applied dc flux across the array changes the peak transparency value from $0.94$ (for the uniform applied flux case) to $0.91$.  This means the transparency is quite robust against noise and disorder. However, simulation shows that increased coupling between meta-atoms reduces the transparency range \cite{Trepanier2015}. 

The observed tunable transparency of the rf SQUID metamaterial offers a range of new previously unattainable functionalities, enabling new applications.  The root cause of the tunable transparency comes from the intrinsic bi-stability of rf-SQUIDs, which can be controlled in a number of ways. Some of these applications capitalize on the effective three-terminal device configuration in which transmission is modulated via a parameter, namely the rf or the dc flux channel.

For example, one can design an rf transmitter where the input rf bias signal modulates the main channel rf transmission. Such a transmitter can be used in quantum computing to shape rf pulses for qubit control \cite{Gambetta2015}, particularly through the scanning direction hysteresis. 

The self-induced transparency modulation feature can also lead to the presently unattainable gain-control capability in SQUID arrays, which are the basis for highly sensitive quantum antennas \cite{Mukhanov2014}. This would require engineering a linear transparency/rf power transfer function in the rf SQUID metamaterial.

Another application is in wideband directly-digitizing rf receivers, in which the rf SQUID metamaterial acts as an input power limiter to protect sensitive superconducting wideband digital-rf receivers from strong jammers \cite{Mukhanov2008}.  With frequency selectivity, one can devise tunable intensity limiting filters to eliminate strong interferors while remaining transparent for other frequencies.

The remarkable sharpness of transparency modulation by dc flux at a fixed applied rf flux amplitude (Fig. \ref{dcflux}) gives an opportunity to use single flux quantum (SFQ) logic to achieve fast transparency switching of the rf-SQUID metamaterial \cite{Trepanier2013, Mukhanov2011review}.  This will enable a range of applications starting from SFQ-modulated  digital communication transmitters to energy-efficient wireless data links between low-power cryogenic SFQ electronics and room-temperature semiconductor modules.  Both of these applications are difficult to solve by means of conventional low-dissipation superconducting electronics.

\section{Conclusion}
In summary, we show that self-induced broad-band transparency is observed in single rf SQUID meta-atoms and in rf SQUID metamaterials. The transparency arises from the bi-stability of individual SQUIDs. The transparency is hysteretic and switchable, and can be tuned by temperature, dc and rf magnetic flux, endowing the metamaterial with an "electromagnetic memory". The transparency range and level can be enhanced through numerous parameters under experimental control. 

\begin{acknowledgments}
This work is supported by the NSF-GOALI and OISE programs through grant $\#$ECCS-1158644, and the Center for Nanophysics and Advanced Materials (CNAM). We thank Masoud Radparvar, Georgy Prokopenko, Jen-Hao Yeh and Tamin Tai for experimental guidance and helpful suggestions, and Alexey Ustinov, Philipp Jung, Susanne Butz, Edward Ott and Thomas Antonsen for helpful discussions. We also thank H. J. Paik and M. V. Moody for use of the pulsed tube refrigerator, as well as Cody Ballard and Rangga Budoyo for use of the dilution refrigerator.
\end{acknowledgments}

\bibliography{bibliography}

\begin{thebibliography}{35}%
\makeatletter
\providecommand \@ifxundefined [1]{%
 \@ifx{#1\undefined}
}%
\providecommand \@ifnum [1]{%
 \ifnum #1\expandafter \@firstoftwo
 \else \expandafter \@secondoftwo
 \fi
}%
\providecommand \@ifx [1]{%
 \ifx #1\expandafter \@firstoftwo
 \else \expandafter \@secondoftwo
 \fi
}%
\providecommand \natexlab [1]{#1}%
\providecommand \enquote  [1]{``#1''}%
\providecommand \bibnamefont  [1]{#1}%
\providecommand \bibfnamefont [1]{#1}%
\providecommand \citenamefont [1]{#1}%
\providecommand \href@noop [0]{\@secondoftwo}%
\providecommand \href [0]{\begingroup \@sanitize@url \@href}%
\providecommand \@href[1]{\@@startlink{#1}\@@href}%
\providecommand \@@href[1]{\endgroup#1\@@endlink}%
\providecommand \@sanitize@url [0]{\catcode `\\12\catcode `\$12\catcode
  `\&12\catcode `\#12\catcode `\^12\catcode `\_12\catcode `\%12\relax}%
\providecommand \@@startlink[1]{}%
\providecommand \@@endlink[0]{}%
\providecommand \url  [0]{\begingroup\@sanitize@url \@url }%
\providecommand \@url [1]{\endgroup\@href {#1}{\urlprefix }}%
\providecommand \urlprefix  [0]{URL }%
\providecommand \Eprint [0]{\href }%
\providecommand \doibase [0]{http://dx.doi.org/}%
\providecommand \selectlanguage [0]{\@gobble}%
\providecommand \bibinfo  [0]{\@secondoftwo}%
\providecommand \bibfield  [0]{\@secondoftwo}%
\providecommand \translation [1]{[#1]}%
\providecommand \BibitemOpen [0]{}%
\providecommand \bibitemStop [0]{}%
\providecommand \bibitemNoStop [0]{.\EOS\space}%
\providecommand \EOS [0]{\spacefactor3000\relax}%
\providecommand \BibitemShut  [1]{\csname bibitem#1\endcsname}%
\let\auto@bib@innerbib\@empty
%</preamble>
\bibitem [{\citenamefont {Harris}\ \emph {et~al.}(1990)\citenamefont {Harris},
  \citenamefont {Field},\ and\ \citenamefont {Imamo\ifmmode~\breve{g}\else
  \u{g}\fi{}lu}}]{Harris1990}%
  \BibitemOpen
  \bibfield  {author} {\bibinfo {author} {\bibfnamefont {S.~E.}\ \bibnamefont
  {Harris}}, \bibinfo {author} {\bibfnamefont {J.~E.}\ \bibnamefont {Field}}, \
  and\ \bibinfo {author} {\bibfnamefont {A.}~\bibnamefont
  {Imamo\ifmmode~\breve{g}\else \u{g}\fi{}lu}},\ }\bibfield  {title} {\enquote
  {\bibinfo {title} {Nonlinear optical processes using electromagnetically
  induced transparency},}\ }\href {\doibase 10.1103/PhysRevLett.64.1107}
  {\bibfield  {journal} {\bibinfo  {journal} {Phys. Rev. Lett.}\ }\textbf
  {\bibinfo {volume} {64}},\ \bibinfo {pages} {1107--1110} (\bibinfo {year}
  {1990})}\BibitemShut {NoStop}%
\bibitem [{\citenamefont {Harris}(1997)}]{Harris1997}%
  \BibitemOpen
  \bibfield  {author} {\bibinfo {author} {\bibfnamefont {S.~H}\ \bibnamefont
  {Harris}},\ }\bibfield  {title} {\enquote {\bibinfo {title}
  {Electromagnetically induced transparency},}\ }\href {\doibase
  10.1063/1.881806} {\bibfield  {journal} {\bibinfo  {journal} {Physics Today}\
  }\textbf {\bibinfo {volume} {50}},\ \bibinfo {pages} {36--42} (\bibinfo
  {year} {1997})}\BibitemShut {NoStop}%
\bibitem [{\citenamefont {Fleischhauer}\ \emph {et~al.}(2005)\citenamefont
  {Fleischhauer}, \citenamefont {Imamoglu},\ and\ \citenamefont
  {Marangos}}]{EITreview3}%
  \BibitemOpen
  \bibfield  {author} {\bibinfo {author} {\bibfnamefont {Michael}\ \bibnamefont
  {Fleischhauer}}, \bibinfo {author} {\bibfnamefont {Atac}\ \bibnamefont
  {Imamoglu}}, \ and\ \bibinfo {author} {\bibfnamefont {Jonathan~P.}\
  \bibnamefont {Marangos}},\ }\bibfield  {title} {\enquote {\bibinfo {title}
  {Electromagnetically induced transparency: Optics in coherent media},}\
  }\href {\doibase 10.1103/RevModPhys.77.633} {\bibfield  {journal} {\bibinfo
  {journal} {Rev. Mod. Phys.}\ }\textbf {\bibinfo {volume} {77}},\ \bibinfo
  {pages} {633--673} (\bibinfo {year} {2005})}\BibitemShut {NoStop}%
\bibitem [{\citenamefont {Fedotov}\ \emph {et~al.}(2007)\citenamefont
  {Fedotov}, \citenamefont {Rose}, \citenamefont {Prosvirnin}, \citenamefont
  {Papasimakis},\ and\ \citenamefont {Zheludev}}]{Fedotov2007}%
  \BibitemOpen
  \bibfield  {author} {\bibinfo {author} {\bibfnamefont {V.~A.}\ \bibnamefont
  {Fedotov}}, \bibinfo {author} {\bibfnamefont {M.}~\bibnamefont {Rose}},
  \bibinfo {author} {\bibfnamefont {S.~L.}\ \bibnamefont {Prosvirnin}},
  \bibinfo {author} {\bibfnamefont {N.}~\bibnamefont {Papasimakis}}, \ and\
  \bibinfo {author} {\bibfnamefont {N.~I.}\ \bibnamefont {Zheludev}},\
  }\bibfield  {title} {\enquote {\bibinfo {title} {Sharp trapped-mode
  resonances in planar metamaterials with a broken structural symmetry},}\
  }\href {\doibase 10.1103/PhysRevLett.99.147401} {\bibfield  {journal}
  {\bibinfo  {journal} {Phys. Rev. Lett.}\ }\textbf {\bibinfo {volume} {99}},\
  \bibinfo {pages} {147401} (\bibinfo {year} {2007})}\BibitemShut {NoStop}%
\bibitem [{\citenamefont {Kurter}\ \emph {et~al.}(2011)\citenamefont {Kurter},
  \citenamefont {Tassin}, \citenamefont {Zhang}, \citenamefont {Koschny},
  \citenamefont {Zhuravel}, \citenamefont {Ustinov}, \citenamefont {Anlage},\
  and\ \citenamefont {Soukoulis}}]{Kurter2011}%
  \BibitemOpen
  \bibfield  {author} {\bibinfo {author} {\bibfnamefont {C.}~\bibnamefont
  {Kurter}}, \bibinfo {author} {\bibfnamefont {P.}~\bibnamefont {Tassin}},
  \bibinfo {author} {\bibfnamefont {L.}~\bibnamefont {Zhang}}, \bibinfo
  {author} {\bibfnamefont {T.}~\bibnamefont {Koschny}}, \bibinfo {author}
  {\bibfnamefont {A.~P.}\ \bibnamefont {Zhuravel}}, \bibinfo {author}
  {\bibfnamefont {A.~V.}\ \bibnamefont {Ustinov}}, \bibinfo {author}
  {\bibfnamefont {S.~M.}\ \bibnamefont {Anlage}}, \ and\ \bibinfo {author}
  {\bibfnamefont {C.~M.}\ \bibnamefont {Soukoulis}},\ }\bibfield  {title}
  {\enquote {\bibinfo {title} {Classical analogue of electromagnetically
  induced transparency with a metal-superconductor hybrid metamaterial},}\
  }\href@noop {} {\bibfield  {journal} {\bibinfo  {journal} {Phys. Rev. Lett.}\
  }\textbf {\bibinfo {volume} {107}},\ \bibinfo {pages} {043901} (\bibinfo
  {year} {2011})}\BibitemShut {NoStop}%
\bibitem [{\citenamefont {Abdumalikov}\ \emph {et~al.}(2010)\citenamefont
  {Abdumalikov}, \citenamefont {Astafiev}, \citenamefont {Zagoskin},
  \citenamefont {Pashkin}, \citenamefont {Nakamura},\ and\ \citenamefont
  {Tsai}}]{Abdumalikov2010}%
  \BibitemOpen
  \bibfield  {author} {\bibinfo {author} {\bibfnamefont {A.~A.}\ \bibnamefont
  {Abdumalikov}}, \bibinfo {author} {\bibfnamefont {O.}~\bibnamefont
  {Astafiev}}, \bibinfo {author} {\bibfnamefont {A.~M.}\ \bibnamefont
  {Zagoskin}}, \bibinfo {author} {\bibfnamefont {Yu.~A.}\ \bibnamefont
  {Pashkin}}, \bibinfo {author} {\bibfnamefont {Y.}~\bibnamefont {Nakamura}}, \
  and\ \bibinfo {author} {\bibfnamefont {J.~S.}\ \bibnamefont {Tsai}},\
  }\bibfield  {title} {\enquote {\bibinfo {title} {Electromagnetically induced
  transparency on a single artificial atom},}\ }\href {\doibase
  10.1103/PhysRevLett.104.193601} {\bibfield  {journal} {\bibinfo  {journal}
  {Phys. Rev. Lett.}\ }\textbf {\bibinfo {volume} {104}},\ \bibinfo {pages}
  {193601} (\bibinfo {year} {2010})}\BibitemShut {NoStop}%
\bibitem [{\citenamefont {Fedotov}\ \emph {et~al.}(2010)\citenamefont
  {Fedotov}, \citenamefont {Tsiatmas}, \citenamefont {Shi}, \citenamefont
  {Buckingham}, \citenamefont {de~Groot}, \citenamefont {Chen}, \citenamefont
  {Wang},\ and\ \citenamefont {Zheludev}}]{Fedotov2010}%
  \BibitemOpen
  \bibfield  {author} {\bibinfo {author} {\bibfnamefont {V.~A.}\ \bibnamefont
  {Fedotov}}, \bibinfo {author} {\bibfnamefont {A.}~\bibnamefont {Tsiatmas}},
  \bibinfo {author} {\bibfnamefont {J.~H.}\ \bibnamefont {Shi}}, \bibinfo
  {author} {\bibfnamefont {R.}~\bibnamefont {Buckingham}}, \bibinfo {author}
  {\bibfnamefont {P.}~\bibnamefont {de~Groot}}, \bibinfo {author}
  {\bibfnamefont {Y.}~\bibnamefont {Chen}}, \bibinfo {author} {\bibfnamefont
  {S.}~\bibnamefont {Wang}}, \ and\ \bibinfo {author} {\bibfnamefont {N.~I.}\
  \bibnamefont {Zheludev}},\ }\bibfield  {title} {\enquote {\bibinfo {title}
  {Temperature control of {F}ano resonances and transmission in superconducting
  metamaterials},}\ }\href@noop {} {\bibfield  {journal} {\bibinfo  {journal}
  {Opt. Express}\ }\textbf {\bibinfo {volume} {18}},\ \bibinfo {pages}
  {9015--9019} (\bibinfo {year} {2010})}\BibitemShut {NoStop}%
\bibitem [{\citenamefont {Jung}\ \emph
  {et~al.}(2014{\natexlab{a}})\citenamefont {Jung}, \citenamefont {Ustinov},\
  and\ \citenamefont {Anlage}}]{Jung2014review}%
  \BibitemOpen
  \bibfield  {author} {\bibinfo {author} {\bibfnamefont {Philipp}\ \bibnamefont
  {Jung}}, \bibinfo {author} {\bibfnamefont {Alexey~V}\ \bibnamefont
  {Ustinov}}, \ and\ \bibinfo {author} {\bibfnamefont {Steven~M}\ \bibnamefont
  {Anlage}},\ }\bibfield  {title} {\enquote {\bibinfo {title} {Progress in
  superconducting metamaterials},}\ }\href
  {http://stacks.iop.org/0953-2048/27/i=7/a=073001} {\bibfield  {journal}
  {\bibinfo  {journal} {Superconductor Science and Technology}\ }\textbf
  {\bibinfo {volume} {27}},\ \bibinfo {pages} {073001} (\bibinfo {year}
  {2014}{\natexlab{a}})}\BibitemShut {NoStop}%
\bibitem [{\citenamefont {Kurter}\ \emph {et~al.}(2012)\citenamefont {Kurter},
  \citenamefont {Tassin}, \citenamefont {Zhuravel}, \citenamefont {Zhang},
  \citenamefont {Koschny}, \citenamefont {Ustinov}, \citenamefont {Soukoulis},\
  and\ \citenamefont {Anlage}}]{Kurter2012}%
  \BibitemOpen
  \bibfield  {author} {\bibinfo {author} {\bibfnamefont {C.}~\bibnamefont
  {Kurter}}, \bibinfo {author} {\bibfnamefont {P.}~\bibnamefont {Tassin}},
  \bibinfo {author} {\bibfnamefont {A.~P.}\ \bibnamefont {Zhuravel}}, \bibinfo
  {author} {\bibfnamefont {L.}~\bibnamefont {Zhang}}, \bibinfo {author}
  {\bibfnamefont {T.}~\bibnamefont {Koschny}}, \bibinfo {author} {\bibfnamefont
  {A.~V.}\ \bibnamefont {Ustinov}}, \bibinfo {author} {\bibfnamefont {C.~M.}\
  \bibnamefont {Soukoulis}}, \ and\ \bibinfo {author} {\bibfnamefont {S.~M.}\
  \bibnamefont {Anlage}},\ }\bibfield  {title} {\enquote {\bibinfo {title}
  {Switching nonlinearity in a superconductor-enhanced metamaterial},}\
  }\href@noop {} {\bibfield  {journal} {\bibinfo  {journal} {Appl. Phys.
  Lett.}\ }\textbf {\bibinfo {volume} {100}},\ \bibinfo {pages} {121906--3}
  (\bibinfo {year} {2012})}\BibitemShut {NoStop}%
\bibitem [{\citenamefont {Tsironis}\ \emph {et~al.}(2014)\citenamefont
  {Tsironis}, \citenamefont {Lazarides},\ and\ \citenamefont
  {Margaris}}]{Tsironis2014}%
  \BibitemOpen
  \bibfield  {author} {\bibinfo {author} {\bibfnamefont {G.P.}\ \bibnamefont
  {Tsironis}}, \bibinfo {author} {\bibfnamefont {N.}~\bibnamefont {Lazarides}},
  \ and\ \bibinfo {author} {\bibfnamefont {I.}~\bibnamefont {Margaris}},\
  }\bibfield  {title} {\enquote {\bibinfo {title} {Wide-band tuneability,
  nonlinear transmission, and dynamic multistability in squid metamaterials},}\
  }\href {\doibase 10.1007/s00339-014-8706-7} {\bibfield  {journal} {\bibinfo
  {journal} {Applied Physics A}\ }\textbf {\bibinfo {volume} {117}},\ \bibinfo
  {pages} {579--588} (\bibinfo {year} {2014})}\BibitemShut {NoStop}%
\bibitem [{\citenamefont {Tinkham}(1996)}]{Tinkham1996}%
  \BibitemOpen
  \bibfield  {author} {\bibinfo {author} {\bibfnamefont {M.}~\bibnamefont
  {Tinkham}},\ }\href@noop {} {\emph {\bibinfo {title} {Introduction to
  Superconductivity}}},\ \bibinfo {edition} {2nd}\ ed.\ (\bibinfo  {publisher}
  {McGraw-Hill},\ \bibinfo {address} {New York},\ \bibinfo {year}
  {1996})\BibitemShut {NoStop}%
\bibitem [{\citenamefont {Anlage}(2011)}]{Anlage2011}%
  \BibitemOpen
  \bibfield  {author} {\bibinfo {author} {\bibfnamefont {S.~M.}\ \bibnamefont
  {Anlage}},\ }\bibfield  {title} {\enquote {\bibinfo {title} {The physics and
  applications of superconducting metamaterials},}\ }\href@noop {} {\bibfield
  {journal} {\bibinfo  {journal} {J. Opt.}\ }\textbf {\bibinfo {volume} {13}},\
  \bibinfo {pages} {024001} (\bibinfo {year} {2011})}\BibitemShut {NoStop}%
\bibitem [{\citenamefont {Lapine}\ \emph {et~al.}(2014)\citenamefont {Lapine},
  \citenamefont {Shadrivov},\ and\ \citenamefont {Kivshar}}]{Lapine2014}%
  \BibitemOpen
  \bibfield  {author} {\bibinfo {author} {\bibfnamefont {Mikhail}\ \bibnamefont
  {Lapine}}, \bibinfo {author} {\bibfnamefont {Ilya~V.}\ \bibnamefont
  {Shadrivov}}, \ and\ \bibinfo {author} {\bibfnamefont {Yuri~S.}\ \bibnamefont
  {Kivshar}},\ }\bibfield  {title} {\enquote {\bibinfo {title}
  {\textit{Colloquium} : Nonlinear metamaterials},}\ }\href {\doibase
  10.1103/RevModPhys.86.1093} {\bibfield  {journal} {\bibinfo  {journal} {Rev.
  Mod. Phys.}\ }\textbf {\bibinfo {volume} {86}},\ \bibinfo {pages}
  {1093--1123} (\bibinfo {year} {2014})}\BibitemShut {NoStop}%
\bibitem [{\citenamefont {Du}\ \emph {et~al.}(2006)\citenamefont {Du},
  \citenamefont {Chen},\ and\ \citenamefont {Li}}]{Du2006}%
  \BibitemOpen
  \bibfield  {author} {\bibinfo {author} {\bibfnamefont {C.~G.}\ \bibnamefont
  {Du}}, \bibinfo {author} {\bibfnamefont {H.~Y.}\ \bibnamefont {Chen}}, \ and\
  \bibinfo {author} {\bibfnamefont {S.~Q.}\ \bibnamefont {Li}},\ }\bibfield
  {title} {\enquote {\bibinfo {title} {Quantum left-handed metamaterial from
  superconducting quantum-interference devices},}\ }\href@noop {} {\bibfield
  {journal} {\bibinfo  {journal} {Phys. Rev. B}\ }\textbf {\bibinfo {volume}
  {74}},\ \bibinfo {pages} {113105} (\bibinfo {year} {2006})}\BibitemShut
  {NoStop}%
\bibitem [{\citenamefont {Lazarides}\ and\ \citenamefont
  {Tsironis}(2007)}]{Lazarides2007}%
  \BibitemOpen
  \bibfield  {author} {\bibinfo {author} {\bibfnamefont {N.}~\bibnamefont
  {Lazarides}}\ and\ \bibinfo {author} {\bibfnamefont {G.~P.}\ \bibnamefont
  {Tsironis}},\ }\bibfield  {title} {\enquote {\bibinfo {title} {{RF}
  superconducting quantum interference device metamaterials},}\ }\href@noop {}
  {\bibfield  {journal} {\bibinfo  {journal} {Appl. Phys. Lett.}\ }\textbf
  {\bibinfo {volume} {90}},\ \bibinfo {pages} {163501} (\bibinfo {year}
  {2007})}\BibitemShut {NoStop}%
\bibitem [{\citenamefont {Maimistov}\ and\ \citenamefont
  {Gabitov}(2010)}]{Maimistov2010}%
  \BibitemOpen
  \bibfield  {author} {\bibinfo {author} {\bibfnamefont {A.~I.}\ \bibnamefont
  {Maimistov}}\ and\ \bibinfo {author} {\bibfnamefont {I.~R.}\ \bibnamefont
  {Gabitov}},\ }\bibfield  {title} {\enquote {\bibinfo {title} {Nonlinear
  response of a thin metamaterial film containing {J}osephson junctions},}\
  }\href@noop {} {\bibfield  {journal} {\bibinfo  {journal} {Opt. Commun.}\
  }\textbf {\bibinfo {volume} {283}},\ \bibinfo {pages} {1633--1639} (\bibinfo
  {year} {2010})}\BibitemShut {NoStop}%
\bibitem [{\citenamefont {Lazarides}\ and\ \citenamefont
  {Tsironis}(2013)}]{Lazarides2013}%
  \BibitemOpen
  \bibfield  {author} {\bibinfo {author} {\bibfnamefont {N.}~\bibnamefont
  {Lazarides}}\ and\ \bibinfo {author} {\bibfnamefont {G.~P.}\ \bibnamefont
  {Tsironis}},\ }\bibfield  {title} {\enquote {\bibinfo {title} {Multistability
  and self-organization in disordered {SQUID} metamaterials},}\ }\href@noop {}
  {\bibfield  {journal} {\bibinfo  {journal} {Supercond. Sci. Technol.}\
  }\textbf {\bibinfo {volume} {26}},\ \bibinfo {pages} {084006} (\bibinfo
  {year} {2013})}\BibitemShut {NoStop}%
\bibitem [{\citenamefont {Jung}\ \emph {et~al.}(2013)\citenamefont {Jung},
  \citenamefont {Butz}, \citenamefont {Shitov},\ and\ \citenamefont
  {Ustinov}}]{Jung2013}%
  \BibitemOpen
  \bibfield  {author} {\bibinfo {author} {\bibfnamefont {P.}~\bibnamefont
  {Jung}}, \bibinfo {author} {\bibfnamefont {S.}~\bibnamefont {Butz}}, \bibinfo
  {author} {\bibfnamefont {S.~V.}\ \bibnamefont {Shitov}}, \ and\ \bibinfo
  {author} {\bibfnamefont {A.~V.}\ \bibnamefont {Ustinov}},\ }\bibfield
  {title} {\enquote {\bibinfo {title} {Low-loss tunable metamaterials using
  superconducting circuits with {J}osephson junctions},}\ }\href@noop {}
  {\bibfield  {journal} {\bibinfo  {journal} {Appl. Phys. Lett.}\ }\textbf
  {\bibinfo {volume} {102}},\ \bibinfo {pages} {062601--4} (\bibinfo {year}
  {2013})}\BibitemShut {NoStop}%
\bibitem [{\citenamefont {Butz}\ \emph {et~al.}(2013)\citenamefont {Butz},
  \citenamefont {Jung}, \citenamefont {Filippenko}, \citenamefont {Koshelets},\
  and\ \citenamefont {Ustinov}}]{Butz20132}%
  \BibitemOpen
  \bibfield  {author} {\bibinfo {author} {\bibfnamefont {S.}~\bibnamefont
  {Butz}}, \bibinfo {author} {\bibfnamefont {P.}~\bibnamefont {Jung}}, \bibinfo
  {author} {\bibfnamefont {L.~V.}\ \bibnamefont {Filippenko}}, \bibinfo
  {author} {\bibfnamefont {V.~P.}\ \bibnamefont {Koshelets}}, \ and\ \bibinfo
  {author} {\bibfnamefont {A.~V.}\ \bibnamefont {Ustinov}},\ }\bibfield
  {title} {\enquote {\bibinfo {title} {A one-dimensional tunable magnetic
  metamaterial},}\ }\href@noop {} {\bibfield  {journal} {\bibinfo  {journal}
  {Opt. Express}\ }\textbf {\bibinfo {volume} {21}},\ \bibinfo {pages}
  {22540--22548} (\bibinfo {year} {2013})}\BibitemShut {NoStop}%
\bibitem [{\citenamefont {Trepanier}\ \emph {et~al.}(2013)\citenamefont
  {Trepanier}, \citenamefont {Zhang}, \citenamefont {Mukhanov},\ and\
  \citenamefont {Anlage}}]{Trepanier2013}%
  \BibitemOpen
  \bibfield  {author} {\bibinfo {author} {\bibfnamefont {M.}~\bibnamefont
  {Trepanier}}, \bibinfo {author} {\bibfnamefont {Daimeng}\ \bibnamefont
  {Zhang}}, \bibinfo {author} {\bibfnamefont {Oleg}\ \bibnamefont {Mukhanov}},
  \ and\ \bibinfo {author} {\bibfnamefont {Steven~M.}\ \bibnamefont {Anlage}},\
  }\bibfield  {title} {\enquote {\bibinfo {title} {Realization and modeling of
  metamaterials made of rf superconducting quantum-interference devices},}\
  }\href {\doibase 10.1103/PhysRevX.3.041029} {\bibfield  {journal} {\bibinfo
  {journal} {Phys. Rev. X}\ }\textbf {\bibinfo {volume} {3}},\ \bibinfo {pages}
  {041029} (\bibinfo {year} {2013})}\BibitemShut {NoStop}%
\bibitem [{\citenamefont {Jung}\ \emph
  {et~al.}(2014{\natexlab{b}})\citenamefont {Jung}, \citenamefont {Butz},
  \citenamefont {Koshelets},\ and\ \citenamefont {Ustinov}}]{Jung2014}%
  \BibitemOpen
  \bibfield  {author} {\bibinfo {author} {\bibfnamefont {P.}~\bibnamefont
  {Jung}}, \bibinfo {author} {\bibfnamefont {M.and Fistul M. V.and
  Lepp\"{a}kangas~J.}\ \bibnamefont {Butz}, \bibfnamefont {S.and~Marthaler}},
  \bibinfo {author} {\bibfnamefont {V.~P.}\ \bibnamefont {Koshelets}}, \ and\
  \bibinfo {author} {\bibfnamefont {A.~V.}\ \bibnamefont {Ustinov}},\
  }\bibfield  {title} {\enquote {\bibinfo {title} {Multistability and switching
  in a superconducting metamaterial},}\ }\href@noop {} {\bibfield  {journal}
  {\bibinfo  {journal} {Nat. Comms.}\ }\textbf {\bibinfo {volume} {5}},\
  \bibinfo {pages} {4730} (\bibinfo {year} {2014}{\natexlab{b}})}\BibitemShut
  {NoStop}%
\bibitem [{\citenamefont {Bishop}\ \emph {et~al.}(2010)\citenamefont {Bishop},
  \citenamefont {Ginossar},\ and\ \citenamefont {Girvin}}]{Bishop2010}%
  \BibitemOpen
  \bibfield  {author} {\bibinfo {author} {\bibfnamefont {Lev~S.}\ \bibnamefont
  {Bishop}}, \bibinfo {author} {\bibfnamefont {Eran}\ \bibnamefont {Ginossar}},
  \ and\ \bibinfo {author} {\bibfnamefont {S.~M.}\ \bibnamefont {Girvin}},\
  }\bibfield  {title} {\enquote {\bibinfo {title} {Response of the strongly
  driven jaynes-cummings oscillator},}\ }\href {\doibase
  10.1103/PhysRevLett.105.100505} {\bibfield  {journal} {\bibinfo  {journal}
  {Phys. Rev. Lett.}\ }\textbf {\bibinfo {volume} {105}},\ \bibinfo {pages}
  {100505} (\bibinfo {year} {2010})}\BibitemShut {NoStop}%
\bibitem [{\citenamefont {Boissonneault}\ \emph {et~al.}(2010)\citenamefont
  {Boissonneault}, \citenamefont {Gambetta},\ and\ \citenamefont
  {Blais}}]{Boissonneault2010}%
  \BibitemOpen
  \bibfield  {author} {\bibinfo {author} {\bibfnamefont {Maxime}\ \bibnamefont
  {Boissonneault}}, \bibinfo {author} {\bibfnamefont {J.~M.}\ \bibnamefont
  {Gambetta}}, \ and\ \bibinfo {author} {\bibfnamefont {Alexandre}\
  \bibnamefont {Blais}},\ }\bibfield  {title} {\enquote {\bibinfo {title}
  {Improved superconducting qubit readout by qubit-induced nonlinearities},}\
  }\href {\doibase 10.1103/PhysRevLett.105.100504} {\bibfield  {journal}
  {\bibinfo  {journal} {Phys. Rev. Lett.}\ }\textbf {\bibinfo {volume} {105}},\
  \bibinfo {pages} {100504} (\bibinfo {year} {2010})}\BibitemShut {NoStop}%
\bibitem [{\citenamefont {Reed}\ \emph {et~al.}(2010)\citenamefont {Reed},
  \citenamefont {DiCarlo}, \citenamefont {Johnson}, \citenamefont {Sun},
  \citenamefont {Schuster}, \citenamefont {Frunzio},\ and\ \citenamefont
  {Schoelkopf}}]{Reed2010}%
  \BibitemOpen
  \bibfield  {author} {\bibinfo {author} {\bibfnamefont {M.~D.}\ \bibnamefont
  {Reed}}, \bibinfo {author} {\bibfnamefont {L.}~\bibnamefont {DiCarlo}},
  \bibinfo {author} {\bibfnamefont {B.~R.}\ \bibnamefont {Johnson}}, \bibinfo
  {author} {\bibfnamefont {L.}~\bibnamefont {Sun}}, \bibinfo {author}
  {\bibfnamefont {D.~I.}\ \bibnamefont {Schuster}}, \bibinfo {author}
  {\bibfnamefont {L.}~\bibnamefont {Frunzio}}, \ and\ \bibinfo {author}
  {\bibfnamefont {R.~J.}\ \bibnamefont {Schoelkopf}},\ }\bibfield  {title}
  {\enquote {\bibinfo {title} {High-fidelity readout in circuit quantum
  electrodynamics using the jaynes-cummings nonlinearity},}\ }\href {\doibase
  10.1103/PhysRevLett.105.173601} {\bibfield  {journal} {\bibinfo  {journal}
  {Phys. Rev. Lett.}\ }\textbf {\bibinfo {volume} {105}},\ \bibinfo {pages}
  {173601} (\bibinfo {year} {2010})}\BibitemShut {NoStop}%
\bibitem [{\citenamefont {Macha}\ \emph {et~al.}(2014)\citenamefont {Macha},
  \citenamefont {Oelsner}, \citenamefont {Reiner}, \citenamefont {Marthaler},
  \citenamefont {Andr\'{e}}, \citenamefont {Sch\"{o}n}, \citenamefont
  {H\"{u}bner}, \citenamefont {Meyer}, \citenamefont {ichev},\ and\
  \citenamefont {Ustinov}}]{Macha2014}%
  \BibitemOpen
  \bibfield  {author} {\bibinfo {author} {\bibfnamefont {Pascal}\ \bibnamefont
  {Macha}}, \bibinfo {author} {\bibfnamefont {Gregor}\ \bibnamefont {Oelsner}},
  \bibinfo {author} {\bibfnamefont {Jan-Michael}\ \bibnamefont {Reiner}},
  \bibinfo {author} {\bibfnamefont {Michael}\ \bibnamefont {Marthaler}},
  \bibinfo {author} {\bibfnamefont {Stephan}\ \bibnamefont {Andr\'{e}}},
  \bibinfo {author} {\bibfnamefont {Gerd}\ \bibnamefont {Sch\"{o}n}}, \bibinfo
  {author} {\bibfnamefont {Uwe}\ \bibnamefont {H\"{u}bner}}, \bibinfo {author}
  {\bibfnamefont {Hans-Georg}\ \bibnamefont {Meyer}}, \bibinfo {author}
  {\bibfnamefont {Evgeni~Il'}\ \bibnamefont {ichev}}, \ and\ \bibinfo {author}
  {\bibfnamefont {Alexey~V.}\ \bibnamefont {Ustinov}},\ }\bibfield  {title}
  {\enquote {\bibinfo {title} {Implementation of a quantum metamaterial using
  superconducting qubits},}\ }\href {\doibase 10.1103/PhysRevLett.105.173601}
  {\bibfield  {journal} {\bibinfo  {journal} {Nat. Comms.}\ }\textbf {\bibinfo
  {volume} {5}},\ \bibinfo {pages} {5146} (\bibinfo {year} {2014})}\BibitemShut
  {NoStop}%
\bibitem [{\citenamefont {Mukhanov}\ \emph {et~al.}(2008)\citenamefont
  {Mukhanov}, \citenamefont {Kirichenko}, \citenamefont {Vernik}, \citenamefont
  {Filippov}, \citenamefont {Kirichenko}, \citenamefont {Webber}, \citenamefont
  {Dotsenko}, \citenamefont {Talalaevskii}, \citenamefont {Tang}, \citenamefont
  {Sahu}, \citenamefont {Shevchenko}, \citenamefont {Miller}, \citenamefont
  {Kaplan}, \citenamefont {Sarwana},\ and\ \citenamefont
  {Gupta}}]{Mukhanov2008}%
  \BibitemOpen
  \bibfield  {author} {\bibinfo {author} {\bibfnamefont {O.~A.}\ \bibnamefont
  {Mukhanov}}, \bibinfo {author} {\bibfnamefont {D.}~\bibnamefont
  {Kirichenko}}, \bibinfo {author} {\bibfnamefont {I.~V.}\ \bibnamefont
  {Vernik}}, \bibinfo {author} {\bibfnamefont {T.~V.}\ \bibnamefont
  {Filippov}}, \bibinfo {author} {\bibfnamefont {A.}~\bibnamefont
  {Kirichenko}}, \bibinfo {author} {\bibfnamefont {R.}~\bibnamefont {Webber}},
  \bibinfo {author} {\bibfnamefont {V.}~\bibnamefont {Dotsenko}}, \bibinfo
  {author} {\bibfnamefont {A.}~\bibnamefont {Talalaevskii}}, \bibinfo {author}
  {\bibfnamefont {J.~C.}\ \bibnamefont {Tang}}, \bibinfo {author}
  {\bibfnamefont {A.}~\bibnamefont {Sahu}}, \bibinfo {author} {\bibfnamefont
  {P.}~\bibnamefont {Shevchenko}}, \bibinfo {author} {\bibfnamefont
  {R.}~\bibnamefont {Miller}}, \bibinfo {author} {\bibfnamefont {S.~B.}\
  \bibnamefont {Kaplan}}, \bibinfo {author} {\bibfnamefont {S.}~\bibnamefont
  {Sarwana}}, \ and\ \bibinfo {author} {\bibfnamefont {D.}~\bibnamefont
  {Gupta}},\ }\bibfield  {title} {\enquote {\bibinfo {title} {Superconductor
  digital-{RF} receiver systems},}\ }\href@noop {} {\bibfield  {journal}
  {\bibinfo  {journal} {IEICE Trans. Electron.}\ }\textbf {\bibinfo {volume}
  {E91-C}},\ \bibinfo {pages} {306--317} (\bibinfo {year} {2008})}\BibitemShut
  {NoStop}%
\bibitem [{\citenamefont {Mukhanov}\ \emph {et~al.}(2014)\citenamefont
  {Mukhanov}, \citenamefont {Prokopenko},\ and\ \citenamefont
  {Romanofsky}}]{Mukhanov2014}%
  \BibitemOpen
  \bibfield  {author} {\bibinfo {author} {\bibfnamefont {O.}~\bibnamefont
  {Mukhanov}}, \bibinfo {author} {\bibfnamefont {G.}~\bibnamefont
  {Prokopenko}}, \ and\ \bibinfo {author} {\bibfnamefont {R.}~\bibnamefont
  {Romanofsky}},\ }\bibfield  {title} {\enquote {\bibinfo {title} {Quantum
  sensitivity: Superconducting quantum interference filter-based microwave
  receivers},}\ }\href {\doibase 10.1109/MMM.2014.2332421} {\bibfield
  {journal} {\bibinfo  {journal} {IEEE Microwave Magazine}\ }\textbf {\bibinfo
  {volume} {15}},\ \bibinfo {pages} {57--65} (\bibinfo {year}
  {2014})}\BibitemShut {NoStop}%
\bibitem [{\citenamefont {Vijay}\ \emph {et~al.}(2009)\citenamefont {Vijay},
  \citenamefont {Devoret},\ and\ \citenamefont {Siddiqi}}]{Vijay2009}%
  \BibitemOpen
  \bibfield  {author} {\bibinfo {author} {\bibfnamefont {R.}~\bibnamefont
  {Vijay}}, \bibinfo {author} {\bibfnamefont {M.~H.}\ \bibnamefont {Devoret}},
  \ and\ \bibinfo {author} {\bibfnamefont {I.}~\bibnamefont {Siddiqi}},\
  }\bibfield  {title} {\enquote {\bibinfo {title} {Invited review article: The
  josephson bifurcation amplifier},}\ }\href {\doibase
  http://dx.doi.org/10.1063/1.3224703} {\bibfield  {journal} {\bibinfo
  {journal} {Review of Scientific Instruments}\ }\textbf {\bibinfo {volume}
  {80}},\ \bibinfo {eid} {111101} (\bibinfo {year} {2009})}\BibitemShut
  {NoStop}%
\bibitem [{\citenamefont {Yaakobi}\ \emph {et~al.}(2013)\citenamefont
  {Yaakobi}, \citenamefont {Friedland}, \citenamefont {Macklin},\ and\
  \citenamefont {Siddiqi}}]{Siddiqi2013}%
  \BibitemOpen
  \bibfield  {author} {\bibinfo {author} {\bibfnamefont {O.}~\bibnamefont
  {Yaakobi}}, \bibinfo {author} {\bibfnamefont {L.}~\bibnamefont {Friedland}},
  \bibinfo {author} {\bibfnamefont {C.}~\bibnamefont {Macklin}}, \ and\
  \bibinfo {author} {\bibfnamefont {I.}~\bibnamefont {Siddiqi}},\ }\bibfield
  {title} {\enquote {\bibinfo {title} {Parametric amplification in josephson
  junction embedded transmission lines},}\ }\href {\doibase
  10.1103/PhysRevB.87.144301} {\bibfield  {journal} {\bibinfo  {journal} {Phys.
  Rev. B}\ }\textbf {\bibinfo {volume} {87}},\ \bibinfo {pages} {144301}
  (\bibinfo {year} {2013})}\BibitemShut {NoStop}%
\bibitem [{\citenamefont {Trepanier}\ \emph {et~al.}(manuscript in
  preparation)\citenamefont {Trepanier}, \citenamefont {Zhang}, \citenamefont
  {Mukhanov},\ and\ \citenamefont {Anlage}}]{Trepanier2015}%
  \BibitemOpen
  \bibfield  {author} {\bibinfo {author} {\bibfnamefont {M.}~\bibnamefont
  {Trepanier}}, \bibinfo {author} {\bibfnamefont {Daimeng}\ \bibnamefont
  {Zhang}}, \bibinfo {author} {\bibfnamefont {Oleg}\ \bibnamefont {Mukhanov}},
  \ and\ \bibinfo {author} {\bibfnamefont {Steven.~M.}\ \bibnamefont
  {Anlage}},\ }\bibfield  {title} {\enquote {\bibinfo {title} {Meta-atom
  interactions and coherent response in superconducting macroscopic quantum
  metamaterials},}\ }\href@noop {} {\  (\bibinfo {year} {manuscript in
  preparation})}\BibitemShut {NoStop}%
\bibitem [{\citenamefont {Cross}\ and\ \citenamefont
  {Gambetta}(2015)}]{Gambetta2015}%
  \BibitemOpen
  \bibfield  {author} {\bibinfo {author} {\bibfnamefont {Andrew~W.}\
  \bibnamefont {Cross}}\ and\ \bibinfo {author} {\bibfnamefont {Jay~M.}\
  \bibnamefont {Gambetta}},\ }\bibfield  {title} {\enquote {\bibinfo {title}
  {Optimized pulse shapes for a resonator-induced phase gate},}\ }\href
  {\doibase 10.1103/PhysRevA.91.032325} {\bibfield  {journal} {\bibinfo
  {journal} {Phys. Rev. A}\ }\textbf {\bibinfo {volume} {91}},\ \bibinfo
  {pages} {032325} (\bibinfo {year} {2015})}\BibitemShut {NoStop}%
\bibitem [{\citenamefont {Mukhanov}(2011)}]{Mukhanov2011review}%
  \BibitemOpen
  \bibfield  {author} {\bibinfo {author} {\bibfnamefont {O.A.}\ \bibnamefont
  {Mukhanov}},\ }\bibfield  {title} {\enquote {\bibinfo {title}
  {Energy-efficient single flux quantum technology},}\ }\href {\doibase
  10.1109/TASC.2010.2096792} {\bibfield  {journal} {\bibinfo  {journal} {IEEE
  Transactions on Applied Superconductivity}\ }\textbf {\bibinfo {volume}
  {21}},\ \bibinfo {pages} {760--769} (\bibinfo {year} {2011})}\BibitemShut
  {NoStop}%
\bibitem [{\citenamefont {Chesca}(1998)}]{Chesca1998}%
  \BibitemOpen
  \bibfield  {author} {\bibinfo {author} {\bibfnamefont {B.}~\bibnamefont
  {Chesca}},\ }\bibfield  {title} {\enquote {\bibinfo {title} {Theory of {RF
  SQUIDs} operating in the presence of large thermal fluctations},}\
  }\href@noop {} {\bibfield  {journal} {\bibinfo  {journal} {J. Low Temp.
  Phys.}\ }\textbf {\bibinfo {volume} {110}},\ \bibinfo {pages} {963--1001}
  (\bibinfo {year} {1998})}\BibitemShut {NoStop}%
\bibitem [{\citenamefont {Likharev}(1986)}]{Likharev1986}%
  \BibitemOpen
  \bibfield  {author} {\bibinfo {author} {\bibfnamefont {K.~K.}\ \bibnamefont
  {Likharev}},\ }\href@noop {} {\emph {\bibinfo {title} {Dynamics of
  {J}osephson Junctions and Circuits}}}\ (\bibinfo  {publisher} {Gordon and
  Breach},\ \bibinfo {address} {New York},\ \bibinfo {year} {1986})\BibitemShut
  {NoStop}%
\bibitem [{\citenamefont {Landau}\ and\ \citenamefont
  {Lifshitz}(1976)}]{Landau1976}%
  \BibitemOpen
  \bibfield  {author} {\bibinfo {author} {\bibfnamefont {L.~D.}\ \bibnamefont
  {Landau}}\ and\ \bibinfo {author} {\bibfnamefont {E.~M.}\ \bibnamefont
  {Lifshitz}},\ }\href@noop {} {\emph {\bibinfo {title} {Mechanics: Volume 1
  (Course of Theoretical Physics S)}}},\ \bibinfo {edition} {3rd}\ ed.\
  (\bibinfo  {publisher} {Butterworth-Heinemann},\ \bibinfo {year}
  {1976})\BibitemShut {NoStop}%
\end{thebibliography}%

\widetext
\newpage

\renewcommand{\baselinestretch}{1.5} 
\begin{center}

\textbf{\large Supplemental Material for Tunable Broadband Transparency of Macroscopic Quantum Superconducting Metamaterials}
\end{center}

{\setstretch{1.5}
\section*{Details of the Experiment}

The sample sits inside a pulsed-tube refrigerator with a base temperature of $4.6$ K. The temperature is controlled via an electric heater connected to a Lake Shore Cryogenic Temperature Controller (Model 340). Our sample is positioned in a rectangular waveguide (either X, Ku, or K-band) so that the rf magnetic field of the propagating wave is perpendicular to the SQUID loop and couples strongly to the meta-atoms (Fig. 1 (a)) \cite{Trepanier2013}. A superconducting coil outside the waveguide provides dc magnetic field， also perpendicular to the SQUID loop. A superconducting wire outside the waveguide can be used to create an intentional dc flux gradient on the sample. The sample is protected from environmental magnetic fields via several layers of magnetic shielding around the waveguide. The transmission and reflection signals from our sample are amplified by a cryogenic low noise amplifier (LNF-LNC6$\_$20A) and a room temperature amplifier (HP 83020A), and are measured by a network analyzer (Agilent N5342A). The dc current through the superconducting coil and the superconducting wire is applied by a Keithley 220 programmable current source.

\section*{Parameters of the samples}

The single rf SQUID meta-atom and the 11x11 array metamaterial were fabricated using the Hypres 0.3 $\mu $A$/ \mu $m$^2$ Nb/AlO$_x$/Nb junction process on silicon substrates, and the meta-atom has a superconducting transition temperature $T_c = 9.2$ K. Two Nb films (135 nm and 300 nm thick) connected by a via and a Josephson junction make up the superconducting loop with geometrical inductance $L$. The capacitance $C$ has two parts: the overlap between two layers of Nb with SiO$_2$ dielectric in between, and the Josephson junction intrinsic capacitance. The rf SQUIDs are designed to be low-noise
($\Gamma = \frac{2 \pi k_B T}{\Phi_0 I_c}<1$ where $T$ is the temperature and $I_c$ is the critical current in the Josephson junction, and
$L_F = \frac{1}{k_B T}\left(\frac{\Phi_0}{2 \pi}\right)^2>> L$ \cite{Chesca1998})
and non-hysteretic ($\beta_{rf}=\frac{2 \pi L I_c}{\Phi_0}<1$).

The parameters for the single meta-atom: geometrical inductance of the rf-SQUID loop $L=280$ pH, the total capacitance $C=0.495$ pF, the geometrical resonant frequency $\omega_{geo}/2\pi=13.52$ GHz, the resistance in the junction $R=1780$ Ohm (4.6 K), $I_c=1.15 \mu$A, $\beta_{rf}=0.98$, rf-SQUID inner diameter $200 \mu$m, outer diameter $800 \mu$m. The $0$ dc flux low-power resonant frequency is at $19$ GHz.

The parameters for meta-atoms of the 11$\times$11 array: $L=55.99$ pH, $C=2.1$ pF, $\omega_{geo}/2\pi=14.85$ GHz, $R=500$ Ohm, $I_c=5 \mu$A, $\beta_{rf}=0.86$, rf-SQUID inner diameter $40 \mu$m, outer diameter $160 \mu$m, center-center distance $170 \mu$m. The $0$ dc flux low-power resonant frequency is at $20.25$ GHz.

\section*{Nonlinear dynamics and numerical simulation}

We treat an rf SQUID as a Resistively and Capacitively Shunted Josephson Junction (RCSJ-model) in parallel with superconducting loop inductance (see Fig. 1 (a)). The macroscopic quantum gauge-invariant phase difference across the junction $\delta$ determines the current through the junction $I=I_c\sin{\delta}$ and the voltage over the junction $V=(\hbar/2e) d\delta / dt$.  The flux quantization condition in an rf-SQUID relates $\delta$ and the total flux $\Phi_{tot}$ through the loop: $\delta = 2 \pi n \Phi_{tot} / \Phi_0$, where $\Phi_0=h/2e$ is the flux quantum and $n$ is any integer. Here we take $n$ to be $1$ because any $2\pi$ added in $\delta$ is essentially an integral constant that can be ignored in the dynamics \cite{Likharev1986}. The total flux through the loop $\Phi_{tot}$ is the combination of applied flux ($\Phi_{dc}+\Phi_{rf}\sin{\omega t}$) and the self-induced flux required to maintain flux quantization:

\renewcommand{\theequation}{S\arabic{equation}}
\begin{equation}
\label{flux}
\Phi_{tot}=\Phi_{dc}+\Phi_{rf} \sin{\omega t}-L(I_c \sin{\delta} + \frac{V}{R} + C\frac{dV}{dt})
\end{equation}

where the term in brackets is the total current through the junction, resistance $R$ and capacitance $C$ in the RCSJ model (see Fig. 1 (a)). Substituting $\Phi_{tot}$ with $\Phi_0 \delta/2\pi $ and $V$ with $(\hbar/2e) d\delta / dt$ into equation (\ref{flux}) and rearranging terms, we arrive at the dimensionless equation:

\begin{equation}
\label{delta}
\frac{d^2\delta}{d\tau^2} + \frac{1}{Q_{geo}} \frac{d\delta}{d\tau} +\delta+ \beta_{rf} \sin{\delta}=(\phi_{dc} + \phi_{rf}\sin{\Omega\tau})
\end{equation}
where $\phi_{dc}=2\pi \Phi_{dc}/\Phi_0$ and $\phi_{rf}= 2\pi \Phi_{rf}/\Phi_0$ are the applied non-dimensional dc flux and rf flux, $\omega_{geo}=\frac{1}{\sqrt{LC}}$ is the geometrical frequency,  $\Omega=\omega/\omega_{geo}$,  $\tau=\omega_{geo}t$, $Q_{geo}=R\sqrt{\frac{C}{L}}$, and $\beta_{rf}=2\pi L I_c/\Phi_0$ is the coefficient determining the degree of nonlinearity in an rf-SQUID. 

An array of $N$ coupled rf-SQUIDs can be described as a system of coupled nonlinear differential equations \cite{Trepanier2015}:
\begin{equation}
\label{array_equation}
 \hat{\delta}+\bar{\bar{\lambda}}(\frac{d^2\hat{\delta}}{d\tau^2} + \frac{1}{Q_{geo}} \frac{d\hat{\delta}}{d\tau} +\hat{\delta}+ \beta_{rf} \sin{\hat{\delta}})=\hat{\phi}_{dc}+\hat{\phi}_{rf}\sin{\Omega \tau}
 \end{equation}
where $\hat{\delta}$ is an $N$-element vector describing the gauge-invariant phases of the $N$ rf-SQUIDs, $\hat{\phi}_{dc}$ and $\hat{\phi}_{rf}$ are $N$-element vectors denoting the non-dimensional dc flux and rf flux applied to each rf-SQUID respectively, $\bar{\bar{\lambda}}$ is an $N$ $\times$ $N$ coupling matrix determining the interactions between the meta-atoms \cite{Trepanier2015}.

We solve equation (\ref{delta}) (or equation (\ref{array_equation}) for an array) numerically to determine $\delta(\tau)$ in steady state. The time-dependent part of $\delta(\tau)$  is very nearly sinusoidal, independent of the drive amplitude. With this, we calculate the dissipated power $V^2/R$ and the effective permeability $\mu_r$ of a meta-atom (or a metamaterial). The transmission and reflection through a partially filled rectangular waveguide with the rf-SQUID medium can be calculated and compared to experiment (Section. \ref{musection}) \cite{Trepanier2013}.

Under a fixed applied dc flux, the normalized transparency value is defined as a function of $\phi_{rf}$ (Fig. 1 and Fig. 2): 
\begin{equation}
\label{transparency_define}
Transparency (\phi_{rf})=1-\frac{|S_{21,res}(\phi_{rf})| (dB)}{|S_{21,res}(0)|(dB)}
\end{equation}
where $|S_{21,res}(\phi_{rf})|$ is the transmission on resonance at a given rf flux, and  $|S_{21,res}(0)|$ is the transmission on resonance when the drive amplitude is low ($\Phi_{rf}/\Phi_{0}<0.001$ for the single rf-SQUID meta-atom). 

When our sample is driven by a fixed rf flux amplitude, the normalized transparency value is a function of the varying dc flux (Fig. 5):
\begin{equation}
\label{transparency_definedc}
Transparency (\phi_{dc})=1-\frac{|S_{21,res}(\phi_{dc})| (dB)}{min(|S_{21,res}(\phi_{dc})|)(dB)}
\end{equation}
where $|S_{21,res}(\phi_{dc})|$ is the transmission on resonance at each dc flux, and  $min(|S_{21,res}(\phi_{dc})|)$ is the minimum transmission on resonance as the dc flux varies. 

\section*{Details of Duffing oscillator approximation}

Given that $\sin{\delta} \simeq \delta-\frac{\delta^3}{3!}$ for small argument, we get the equation of a single rf-SQUID in the Duffing oscillator approximation, which can be solved analytically: 
\begin{equation}
\label{Duffing_1}
\frac{d^2\delta}{d\tau^2}+ \frac{1}{Q_{geo}}\frac{d\delta}{d\tau} + (1+\beta_{rf}) \delta -\frac{\beta_{rf}}{6}\delta^3 =\phi_{dc} + \phi_{rf}\sin{\Omega \tau}
\end{equation}

In the Duffing oscillator approximation, the natural frequency of the oscillating $\delta$ when $\phi_{dc}=0$ is $\omega_0=\sqrt{1+\beta_{rf}}\omega_{geo}$; it determines the resonant frequency of an rf-SQUID when the drive amplitude is very low. For the single meta-atom, $\omega_0/2\pi=19$ GHz.  The $\beta_{rf}/6$ prefactor of the nonlinear term $\delta^3$ will modify the resonance properties when the drive amplitude is high. 

We separate the dc and ac parts of $\delta (t)$ for analysis. With the ansatz $\delta=\delta_{dc}+\delta_{rf}$ in equation (\ref{Duffing_1}), we get two coupled equations for $\delta_{dc}$ and $\delta_{rf}$.

\begin{equation}
\label{deltadc1}
-\frac{\beta_{rf}}{6}\delta_{dc}^3+(1+\beta_{rf})\delta_{dc}-\phi_{dc}=0
\end{equation}

\begin{equation}
\label{deltarf}
\begin{multlined}
\frac{d^2\delta_{rf}}{d\tau^2}+ \frac{1}{Q_{geo}}\frac{d\delta_{rf}}{d\tau} + (1+\beta_{rf}-\frac{\beta_{rf}}{2}\delta_{dc}^2) \delta_{rf} \\-
\frac{\beta_{rf}}{6}\delta_{rf}^3-\frac{\beta_{rf}}{2}\delta_{dc}\delta_{rf}^2=\phi_{rf}\sin{\Omega \tau} 
\end{multlined}
\end{equation}

The non-zero $\phi_{dc}$ gives rise to a non-zero $\delta_{dc}$, which adds a $\delta_{rf}^2$ term in the ac gauge invariant phase oscillation.  Also, the resonant frequency at low drive with a non-zero dc flux is modified to $\omega_1=\omega_{geo}\sqrt{1+\beta_{rf}-(\beta_{rf}/2)\delta_{dc}^2}$ (smaller than $\omega_0$).

If we make the ansatz that $\delta_{rf}=b\sin{(\Omega \tau+\alpha)}$  \cite{Landau1976}, one finds a cubic equation for the amplitude $b^2$,
\begin{equation}
\label{cubic}
b^2[(\epsilon-\kappa b^2)^2+(\frac{\omega_{geo}}{2Q_{geo}})^2]=[\frac{(\omega_{geo}^2 \phi_{rf})^2 }{2\omega_1}]^2
\end{equation}
where $\epsilon=\omega-\omega_{1}$ is the difference between driving frequency and the natural frequency, and $\kappa$ is the anharmonicity coefficient given by: 

\begin{equation}
\label{kappa}
\kappa=-\frac{\beta_{rf}\omega_{geo}^2}{16\omega_1}(1+\frac{5}{3}\frac{ \beta_{rf} \omega_{geo}^2 \delta_{dc}^2}{\omega_1^2})
\end{equation}

The number of real roots of equation (\ref{cubic}) changes with driving frequency and driving amplitude. When the driving amplitude is very small, equation (\ref{cubic}) has one real root for $b^2$ throughout the whole frequency range. The peak value of $b$ denotes resonance. As driving amplitude of the meta-atom increases, the resonance bends towards the lower frequency side, but is still single-valued. After the driving amplitude reaches a critical value $\phi_{rf,bi}$ determined by
\begin{equation}
\label{criticalfrf}
\phi_{rf,bi}=2 \sqrt{\frac{\omega_1^2}{3\sqrt{3}|\kappa|Q_{geo}^3\omega_{geo}}},
\end{equation}
equation (\ref{cubic}) has three real roots for $b^2$ in a range of frequencies (Fig. 3 (d)). The middle-value root is unstable, so finally we get bi-stability in a certain range of frequencies at a high enough driving amplitude. 

With a non-zero dc flux, the oscillating $\delta$ now has a dc part, and the natural frequency decreases to $\omega_1$, which further modifies the anharmonicity coefficient $\kappa$. Since the onset of transparency predicted by Eq. \ref{criticalfrf} is related the resonant frequency $\omega_1$ and $\kappa$, we can tune the onset of transparency by dc flux. For example, with a dc flux of a quarter flux quantum, the single-SQUID has an onset $\phi_{rf,bi}/2\pi$ that is $13\%$ smaller than the $0$-flux case. 

\section*{Effective Permeability}
\label{musection}

The magnetic susceptibility of the rf-SQUID metamaterial can be calculated as $<\Phi_{ac}/\Phi_{rf}\sin{\Omega\tau}>$, where $\Phi_{ac}$ is the time dependent part of the flux induced in the SQUID. Further we can calculate effective magnetic permeability by knowing the filling factor $F$ of the metamaterial inside a waveguide. 
\begin{equation}
\label{mu}
\mu_r=1+F\left(\left<\frac{\Phi_{ac}}{\Phi_{rf}\sin{\Omega \tau}}\right>-1\right),
\end{equation}
%We calculate $\mu_r$ as a function of frequency and rf flux in the transparent state of the rf-SQUID metamaterial. For each rf flux value, we find the minimum value of the real part (blue dot line) in rhe frequency domain and the maximum value of the imaginary part (red dashed line) (Fig. S\ref{mu_trans} ). In the transparent regime the real part of $\mu_r$ approaches $1.0$ and the imaginary part reaches $0.0$, indicating that the metamaterial can be manipulated by rf flux to become magnetically inert. For a filling factor of only $8\%$, the metamaterial already shows a large negative effective $\mu_r$.

We calculate $\mu_r$ as a function of frequency as the drive amplitude through the rf-SQUID metamaterial varies (Fig. \ref{mu_freq} ). In the transparent regime the real part of $\mu_r$ approaches $1.0$ and the imaginary part reaches $0.0$, indicating that the metamaterial can be manipulated by rf flux to become magnetically inert. The steps in $\Phi_{rf}/\Phi_0=0.3$ case indicate the multi-stability at high drive amplitude predicted in previous work \cite{Jung2014}. For a filling factor of only $8\%$, the metamaterial already shows a large negative effective $\mu_r$.

The transmission through a partially filled rectangular waveguide can be calculated as
\begin{equation}
S_{21}=\frac{\sqrt{\gamma}}{\cos{kl}-\frac{i}{2}\left(\frac{1}{\gamma}+\gamma\right)\sin{kl}},
\label{effectivemed}
\end{equation}
where $l$ is the effective length of the medium, $k=\sqrt{\mu_r\left(\frac{\omega}{c}\right)^2-\left(\frac{\pi}{a}\right)^2}$ is the wave number in the medium, $k_0=\sqrt{\left(\frac{\omega}{c}\right)^2-\left(\frac{\pi}{a}\right)^2}$ is the wave number in the empty waveguide, $c$ is the speed of light, $a$ is the waveguide dimension, and $\gamma=\frac{k}{\mu_r k_0}$. Conversely, the effective $\mu_r$ can be extracted from measured complex transmission data by employing equation (\ref{effectivemed}). The $\mu_r$ extracted from experimental data shows the magnetically inert behavior in transparent regime as well. 

\renewcommand{\thefigure}{S\arabic{figure}}

\setcounter{figure}{0}

\begin{figure}[]
\centering
\includegraphics[width=160 mm]{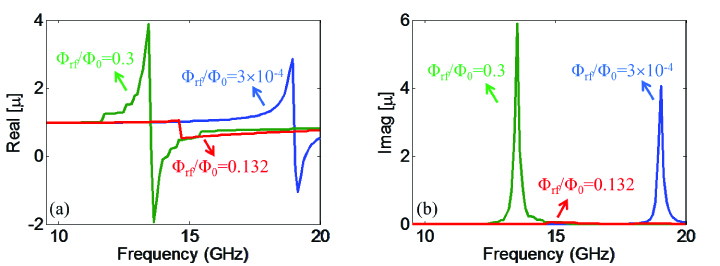}
\caption{The (a) real part $Real[\mu_r]$ and (b) imaginary part $Imag[\mu_r]$ of the effective magnetic permeability $\mu_r$ of an rf-SQUID metamaterial in simulation. The effective $\mu_r$ as a function of frequency is plotted for input rf flux $\Phi_{rf}/\Phi_0=3\times10^{-4}$ (Blue line), $\Phi_{rf}/\Phi_0=0.132$ (Red line), $\Phi_{rf}/\Phi_0=0.3$ (Green line). The parameters: $L=280$ pH, $C=0.49$ pF, $R=1780$ Ohm, $F=8\%$, rf-SQUID inner diameter: $100$ $\mu$m, outer diameter: $800$ $\mu$m.}
\label{mu_freq}
\end{figure}

\section*{Transparency Range Tuning}
We evaluate the onset $\Phi_{rf}/\Phi_{0}$ of transparency by the Duffing oscillator approximation, and the upper critical $\Phi_{rf}/\Phi_{0}$ by methods discussed in \cite{Jung2014}. The two limits of transparency can be tuned by different parameters (Fig. \ref{trans_range}). Larger critical current $I_c$ increases significantly the upper critical $\Phi_{rf}/\Phi_0$, while the higher capacitance and resistance reduce the onset $\Phi_{rf}/\Phi_0$ for transparency. The transparencty range greatly broadens when either of these three parameters is increased. At lower temperature, the critical current and resistance increase simultaneously, resulting in a wider transparency $\Phi_{rf}/\Phi_0$ range. The enhancement of transparency range at lower temperature is consistent with both experimental and numerical results.

\begin{figure}[]
\centering
\includegraphics[width=80 mm]{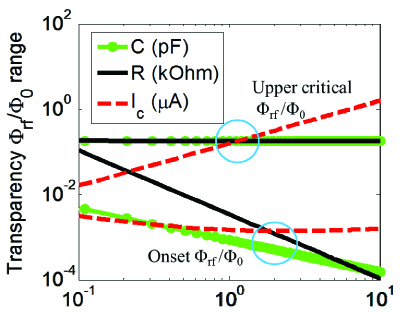}
\caption{The transparency rf flux range tuned by capacitance (green line, $R=1780$ ohm, $I_c=1.15$ $\mu$A), resistance in the junction (black solid line, $C=0.49$ pF, $I_c=1.15$ $\mu$A), and the critical current (red dashed line, $R=1780$ ohm, $C=0.49$ pF). In all cases $L=280$ pH, rf-SQUID inner diameter: $100$ $\mu$m, outer diameter: $800$ $\mu$m.}
\label{trans_range}
\end{figure}

\section*{Bi-stability of fold-over resonance}
In the bi-stability regime, a frequency scan and rf flux amplitude scan both result in a hysteretic transparency behavior. The fold-over resonance (Fig. \ref{foldover} (a)) predicted in the Duffing oscillator approximation shows that %the frequency scan gives a transparent state when sweeping from low to high values, and the reverse frequency scan excites the dissipative resonating state.%
if the driving frequency scans from low to high, the amplitude of $\delta$ follows the trace F-E-D-B-A, where the largest amplitude is at point B, denoting the transparent resonance state ($\delta_{TR}$). A reversed frequency scan of trace A-B-C-E-F makes point $C$ the dissipative resonance ($\delta_{OP}$).

The hysteresis in rf flux is similar. Fig. \ref{foldover} (b) plots the bi-stable fold-over resonance of an rf-SQUID meta-atom at drive amplitude $\Phi_{rf}/\Phi_0=0.003$ (red curve) and $\Phi_{rf}/\Phi_0=0.006$ (black curve).
As the rf flux amplitude keeps increasing at a fixed frequency, in this case $17.6$ GHz, the amplitude of $\delta$ changes from $A_1$ to $B_1$, and stays in the transparent state. However, if the scan starts from very high rf flux value and gradually decreases, the higher amplitude $B_1'$ which denotes the dissipative mode will be excited. Further reduced rf flux $\Phi_{rf}/\Phi_0=0.003$ brings the rf-SQUID back to the $A_1$ state. The highest amplitude at a fixed frequency ($17.6$ GHz) is at $B_1$ when sweeping rf flux from low to high values, while it is at $B_1'$ for a reverse scan. The same applies for the rf flux amplitude scan at $18.5$ GHz: a forward sweep in rf flux amplitude keeps the amplitude of $\delta$ in the lower values ($A_2$ to $B_2$), while the backward sweep excites $\delta$ to oscillate at $B_2'$ and $A_2'$, resulting in a dissipative state.
 
\begin{figure}[]
\centering
\includegraphics[width=160 mm]{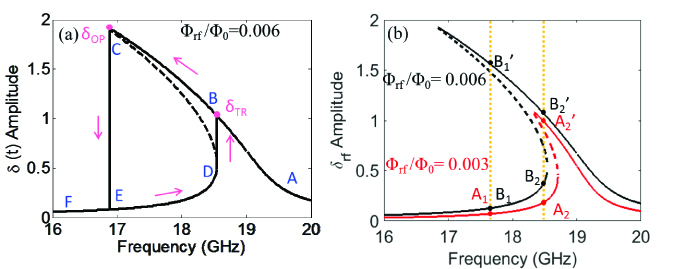}
\caption{The analytically calculated fold-over resonance of an rf-SQUID meta-atom in the Duffing approximation. (a) The fold-over resonance when $\Phi_{rf}/\Phi_0=0.006$. The region between B-C-E-D is hysteretic. Depending on scanning direction, the resonance can either happen at point B with amplitude of $\delta_{TR}$ or at point C with amplitude of $\delta_{OP}$. (b) The calculated fold-over resonances for two rf flux values: $\Phi_{rf}/\Phi_0=0.003$ (red line), and $\Phi_{rf}/\Phi_0=0.006$ (black line). Dashed lines denote the unstable state. $A_1$ ($B_1$) and $A_2$ ($B_2$) are the lower-amplitude roots in the bi-stability regime for $\Phi_{rf}/\Phi_0=0.003$ ($\Phi_{rf}/\Phi_0=0.006$) case. $A_2'$ is the higher-amplitude root in the bi-stability regime for $\Phi_{rf}/\Phi_0=0.003$ at the same frequency of $A_2$. $B_1'$ and $B_2'$ are the higher-amplitude roots in the bi-stability regime for $\Phi_{rf}/\Phi_0=0.006$.}
\label{foldover}
\end{figure}

\section*{Pulsed-rf measurement}
In Fig. 1 in the main text, the experiment and simulation match well except that the upper critical rf flux differs by about $0.12 \Phi_0$. We believe the difference is due to the local heat generated on our sample from high input power that turns the transparency state to the dissipative phase slip state prematurely. To reduce the local heating, we conducted a series of pulsed-rf measurements, where the input signal is periodically pulsed with the pulse width long enough so that $\delta (t)$ achieves a steady state. Pulsed rf measurements give larger transparency and better resemblance to simulation. For similar reasons, the experimental data in Fig. 4 (a) in the main text was taken with the pulsed rf measurement with a duty cycle of $1\%$.

}

%\bibliography{bibliography}

\end{document}